\documentclass{PoS}

\usepackage{dcolumn}
\usepackage{bm}
\usepackage{graphicx}
\usepackage{epstopdf}
\usepackage{epsfig}
\usepackage{color}
\usepackage{wasysym}
\usepackage{twoopt}
\usepackage{verbatim}
\usepackage{dashrule}
\newcommand{\He}{\textsf{He}}
\newcommand{\Li}{\textsf{Li}}
\newcommand{\Be}{\textsf{Be}}
\newcommand{\B}{\textsf{B}}
\newcommand{\C}{\textsf{C}}
\newcommand{\N}{\textsf{N}}
\newcommand{\Oxy}{\textsf{O}}

\newcommand{\Fe}{\textsf{Fe}}
  
\newcommand{\BC}{\textsf{B}/\textsf{C}}

\newcommand{\epfrac}{e\ensuremath{^{+}}/(e\ensuremath{^{-}}\,+\,e\ensuremath{^{+}})} 
\newcommand{\epm}{\ensuremath{e^{\pm}}}
\newcommand{\pbarp}{\textsf{\ensuremath{\bar{p}/p}}}

\newcommand{\Dragon}{\texttt{DRAGON}}

\newcommand{\ApJ}{ApJ}
\newcommand{\AeA}{A\&A}
\newcommand{\PRL}{PRL}
\newcommand{\PRD}{PRD}
\newcommand{\JCAP}{JCAP}
\newcommand{\PLB}{PLB}
\newcommand{\etal}{et alii}

\newcommand{\AMS}{\textsf{AMS}}
 
\newcommand{\ie}{\textit{i.e.}}

\def\citep#1{{\cite{#1}}}
\def\citet#1{{\cite{#1}}}

\title{Inhomogeneous diffusion model for recent data on high-energy cosmic rays}
\ShortTitle{Inhomogeneous diffusive transport models for Galactic CRs}

\author{\speaker{Nicola Tomassetti}\\ %\thanks{A footnote may follow.}\\
LPSC, Universit\'e Grenoble-Alpes, CNRS/IN2P3, F-38026 Grenoble, France; email: nicola.tomassetti@lpsc.in2p3.ch\\
        E-mail: \email{nicola.tomassetti@lpsc.in2p3.fr}}

\abstract{
The \AMS{} Collaboration has recently released precision data on cosmic ray (CR) leptons and protons at high energies. Interesting progresses have also been made on the measurement of CR nuclei, such as the boron-to-carbon ratio or the lithium spectrum, up to TeV/nucleon energies. In order to provide a description these data, I consider a diffusion model of CR propagation which allows for latitudinal variations of the CR diffusion properties in the Galactic halo. I discuss the role of high-precision data on light CR nuclei in resolutely testing this model and the key propagation parameters.
}

\FullConference{The 34th International Cosmic Ray Conference,\\
		30 July- 6 August, 2015\\
		The Hague, The Netherlands}

\begin{document}

%%%%%%%%%%%%%%%%%%%%%%%%%%%%
\section{Introduction}  %%%%
%%%%%%%%%%%%%%%%%%%%%%%%%%%%

The observed properties of Galactic cosmic rays (CRs) are believed to
arise from a combination of two basic plasma astrophysics phenomena: %, namely, 
diffusive shock acceleration mechanisms, occurring in supernova remnants, 
and diffusive trasport processes off magnetic turbulence \citep{Bell2014,Blasi2013,Grenier2015}. 
The paradigm of CR transport as a diffusion process off magnetic irregularities has been established for decades.
It may explain the observed high-degree of isotropy on the CR arrival directions or the abundance of
the secondary \Li-\Be-\B{} elements that are generated from CR collisions with the matter. 
The so-called \emph{standard models} based on this understanding employ several simplifying assumptions. 
In practical implementations, the CRs are assumed confined inside a cylindrical halo, that encompasses the Galactic plane,
after being released by supernova remnants with power-law acceleration spectra $Q\propto E^{-\nu}$. 
The diffusion coefficient is assumed to be $K(E)\propto E^{\delta}$,
spatially homogeneous and isotropic in the whole halo. 
The combined effects of acceleration and transport leads to power-law spectra $\phi(E)\sim Q/K \propto E^{-\delta-\nu}$ for primary 
components (protons, \He, \C, \Oxy, \Fe) and a dependence $\sim E^{-\delta}$ for secondary/primary ratios.
The data on primary spectra and on the \BC{} ratio constrain the parameters $\delta\approx$\,0.3--0.7 and $\nu\approx$\,2--2.4 \citep{Strong2007}.

Recent experiments such as \AMS{} PAMELA or CREAM have found exciting features
in the CR spectrum that may suggest the need of a revision of some standard model assumptions.
The rise of the positron fraction \epfrac{} at $\sim$10--300 GeV of energy \citep{Accardo2014,Adriani2009}, in contrast
to the standard expectations, may suggest the presence of dark-matter or astrophysical sources of high-energy $e^{\pm}$,
as well as the need to reassess the secondary \epm{} secondary production \citep{Choi2014,Blum2013,Serpico2011,Blasi2009,TomassettiDonato2015}. 
The spectral hardening observed in CR protons and nuclei at $E\sim$\,300\,GeV/nucleon \citep{Adriani2011,Yoon2011,Panov2009,Aguilar2015}
may be a signature of new astrophysical phenomena, occurring either in acceleration or 
in propagation, that are not accounted by the standard models \citep{Ptuskin2013,Tomassetti2012,Blasi2012,Vladimirov2012,ThoudamHorandel2014}.
Revisiting CR propagation is particular suggestive given its connection with the so-called ``anisotropy problem''.
The data on CR anisotropy amplitude prefers a rather shallow dependence of the high-energy diffusivity
in comparison with that inferred from the \BC{} ratio \citep{Pohl2013}.
A change on diffusion is also hinted by a high-energy flattening on the secondary to primary ratios.
New \pbarp{} data presented by \AMS{} have generated widespread interest in connection with
for the dark matter searches or with the astrophysical diagnostics of CR propagation \citep{Giesen2015,Kappl2015,Cowsik2015}.

In the following, we present a numerical calculation of CR spectra under a \emph{two halo} scenario of inhomogeneous diffusive transport. 
In this scenario, we assume that the diffusion is shallower when CRs propagate close to the Galactic disk,
based on the idea that the two regions are characterized by a different of magnetic turbulence regimes.
This scenario represents the simplest but physically consistent generalization of the standard models
that generally employ a homogeneous diffusivity in the whole propagation region.
In standard models, even when a spatial dependence of CR diffusion is considered, the energy dependence 
of the diffusion coefficient is always assumed to be ``universal'', \ie, 
standard models assume a unique diffusion regime in the whole Galaxy.
On the contrary, we believe that the different types of turbulence observed in the Galaxy lead to
different \emph{energy dependence} of CR diffusion. 
By the mathematical point of view, this suggestion is formalized by a spatial dependent 
diffusion coefficient which is \emph{not separable into energy and space coordinates}.
As we will show, the two-halo scenario of CR diffusion is supported by several CR data, with
important consequences for the physics observables that are being investigated by \AMS. 
In particular, we will comment its implications for dark matter searches via antimatter.

%%%%%%%%%%%%%%%%%%%%%%%%%
\section{The model}  %%%%
%%%%%%%%%%%%%%%%%%%%%%%%%

We define a phenomenological scenario with two propagation regions, \ie, two \emph{halos}, characterized
by different energy dependence of the diffusion coefficient.
The inner halo represents a thin region, surrounding the disk by a few hundred pc, 
where the turbulence of the medium is influenced by supernova explosions. 
The outer halo represents a wider region, extended up to $z=L$, where the turbulence 
is presumably generated by CRs themselves, given that supernova explosions do not occurs away from the disk.
To numerically set the model, we employ the \Dragon{} package \citep{Dragon}, which
solves the diffusion equation for a given distribution of sources, providing
equilibrium solutions for the interstellar spectra of nuclei and leptons. 
We introduced a modification on the finite-differencing scheme and on the spatial grid
in order to have a latitudinal dependence for the parameter $\delta\equiv\delta(z)$,
The implementation is close to the work in \citep{Jin2015}. 
The vertical extent of the full propagation region is $L=$\,4\,kpc. The two halos are sized 
$l_{i}=\xi L$ and $l_{o}=(1-\xi)L$, with $0<\xi<1$. 
The diffusion coefficient is $K(R,z)= \beta K_{0}\left( R/R_{0}\right)^{\delta(z)}$.
With this functional form, the diffusion coefficient $K$ is non-separable into energy and space coordinates.
Its spatial dependence is defined by $\delta=\delta_{i}$ in the inner halo ($|z|< \xi L$) and $\delta=\delta_{o}\equiv\delta_{i}+\Delta$ 
in the outer halo ($\xi L < |z| < L$). The two $\delta_{i/o}$ values are connected smoothly
in order to ensure the continuity of $K(z)$ across the two zones. 
From the consideration we made, one has $\delta_{i}<\delta_{o}$, \ie, $\Delta>0$.
In the following, we adopt the parameters $\delta=$\,0.15, $\Delta=$\,0.5, and $\xi=$\,0.14. 
\begin{figure}[!t] 
  \includegraphics[width=0.45\textwidth]{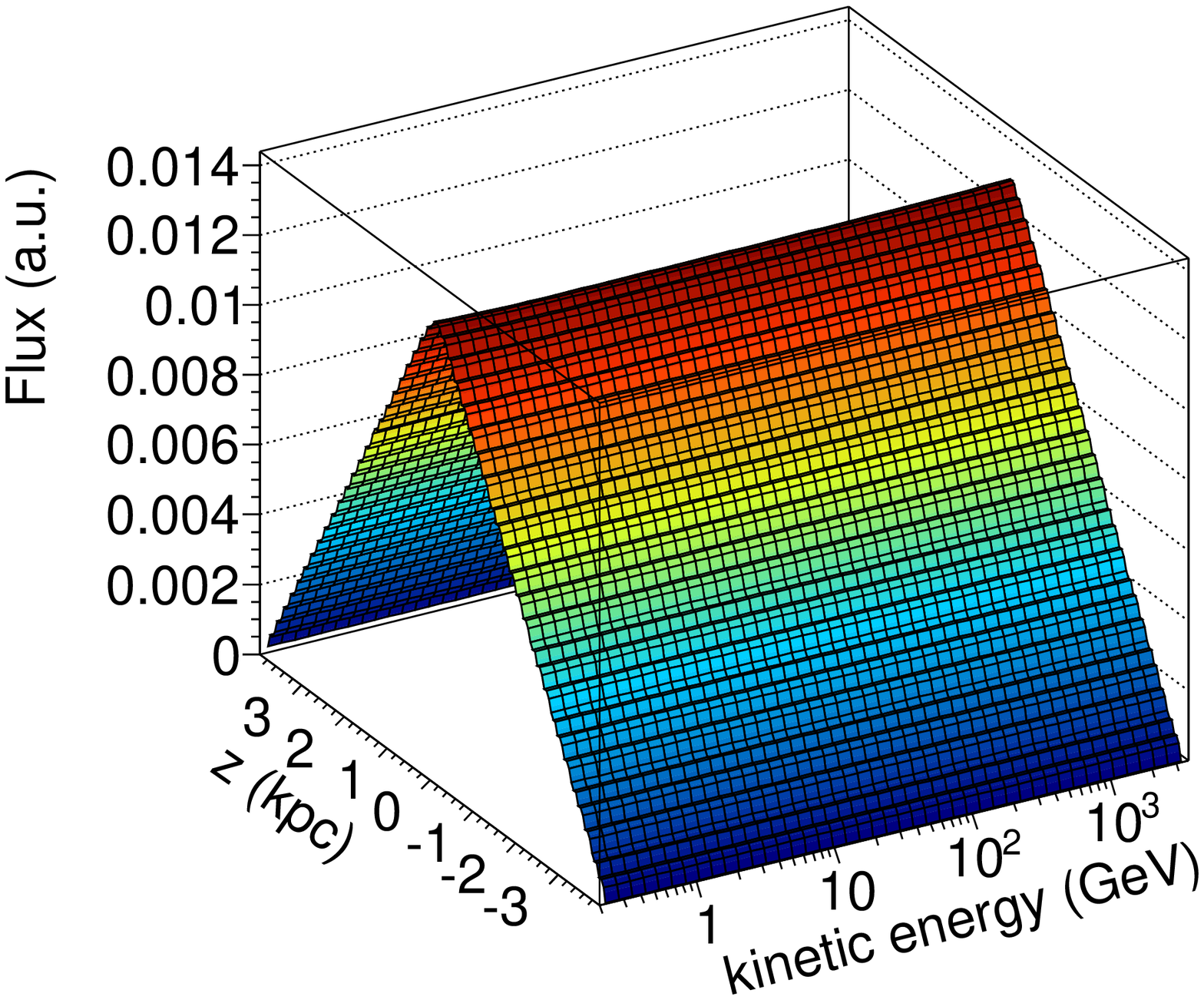}
  \qquad
  \includegraphics[width=0.45\textwidth]{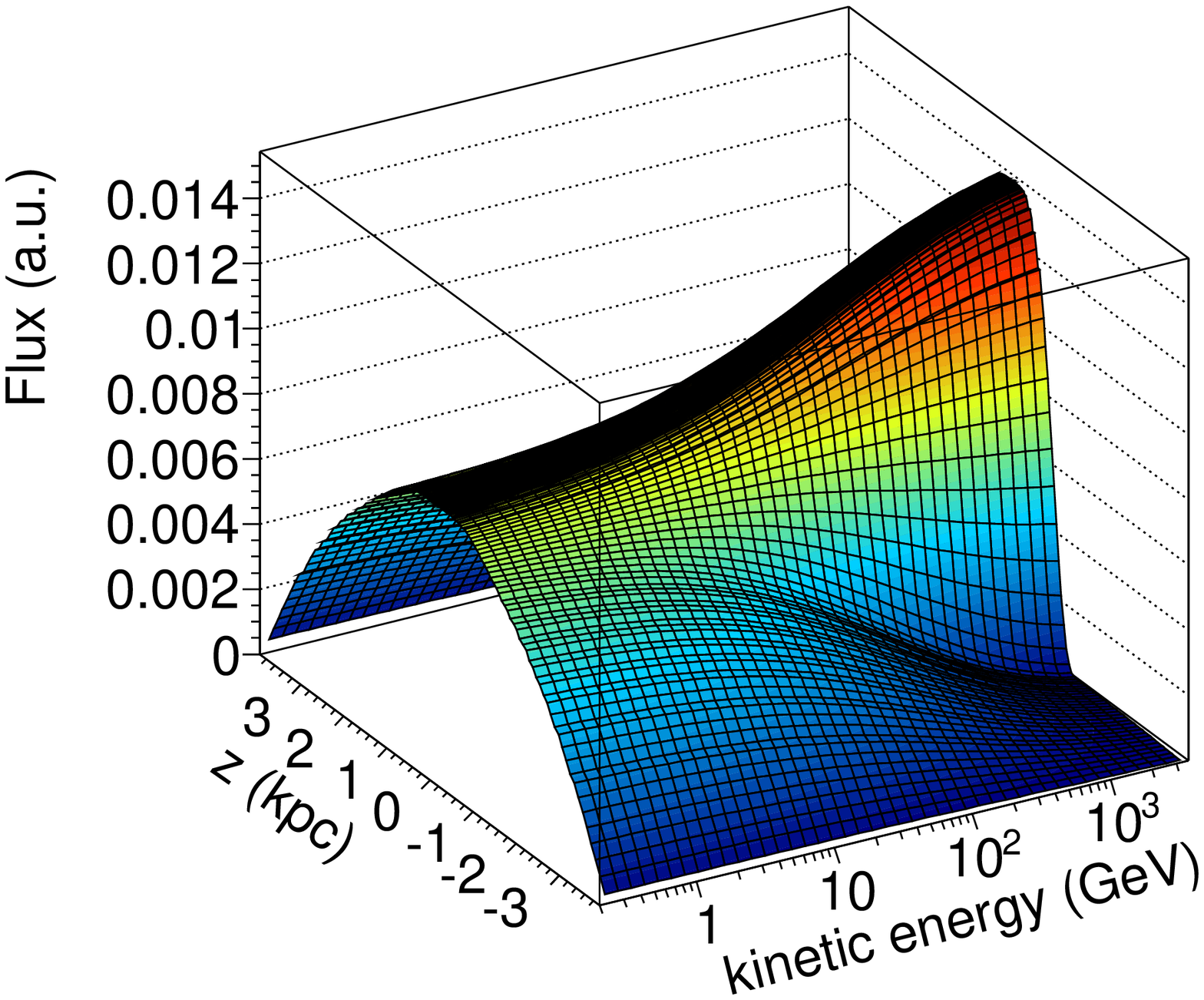}
\caption{\footnotesize%
  CR flux density for primary protons as function of energy and the spatial $z$-coordinate 
  for the standard diffusion model (left) and from the inhomogeneous model (right).
}\label{Fig::ccProtonDensity}
\end{figure}
The source spectra are of the type $Q(R)=\beta\left(R/R_{0}\right)^{-\nu}$ with index $\nu$ tuned to match the data.
In this implementation, the nuclear physics inputs such as fragmentation, decays, energy losses are from \textsf{Galprop} code \citep{Strong2007}.
We also set up a \emph{standard model} of Kraichnan-type diffusion, 
with $\delta\equiv$\,0.5 everywhere in the whole halo,
The solar modulation is described under the \emph{force-field} approximation \citep{Gleeson1968}.

%%%%%%%%%%%%%%%%%%%%%%%%%%%%%%%%%%%%%%%%%%%%%%%%%%%%%%%%
\section{Predictions for Primary Cosmic-Ray Hadrons}  %%%
%%%%%%%%%%%%%%%%%%%%%%%%%%%%%%%%%%%%%%%%%%%%%%%%%%%%%%%%

The flux density as function of energy and of the spatial $z$--coordinate is shown in Fig.\,\ref{Fig::ccProtonDensity}
for the proton spectrum. Other species have similar shapes. 
The particular shape of the two-halo model reflect the interplay of the CR diffusion into the two propagation halos. 
For the equilibrium fluxes expected at Earth, this interplay produces a remarkable 
departure from the standard power-law expectations. 
%%%%%%%%%%%%%   PROTON SPECTRUM %%%%%%%%%%%%%%%%%%%%
\begin{figure}[!t]
\includegraphics[width=0.46\textwidth]{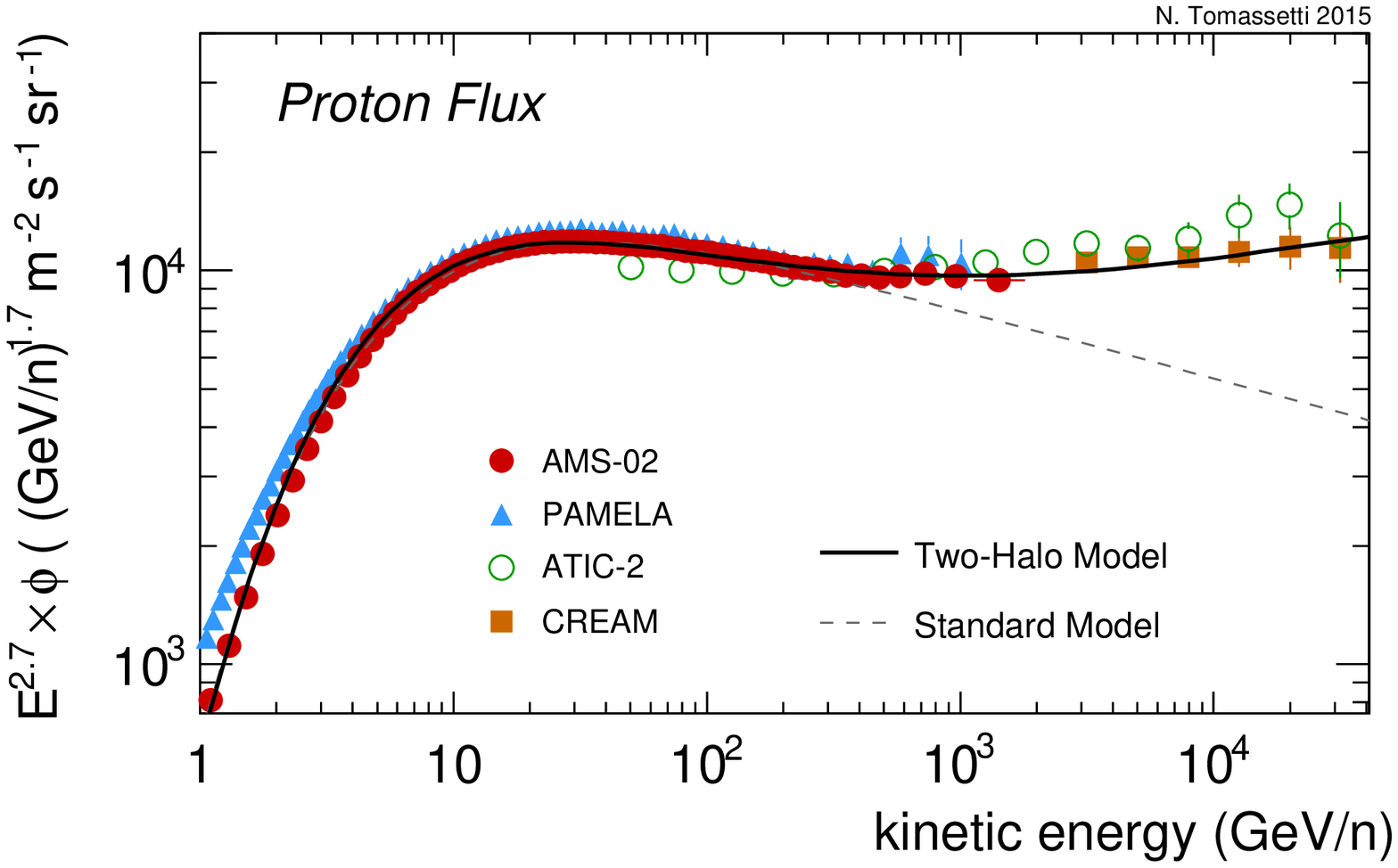}
\qquad
\includegraphics[width=0.46\textwidth]{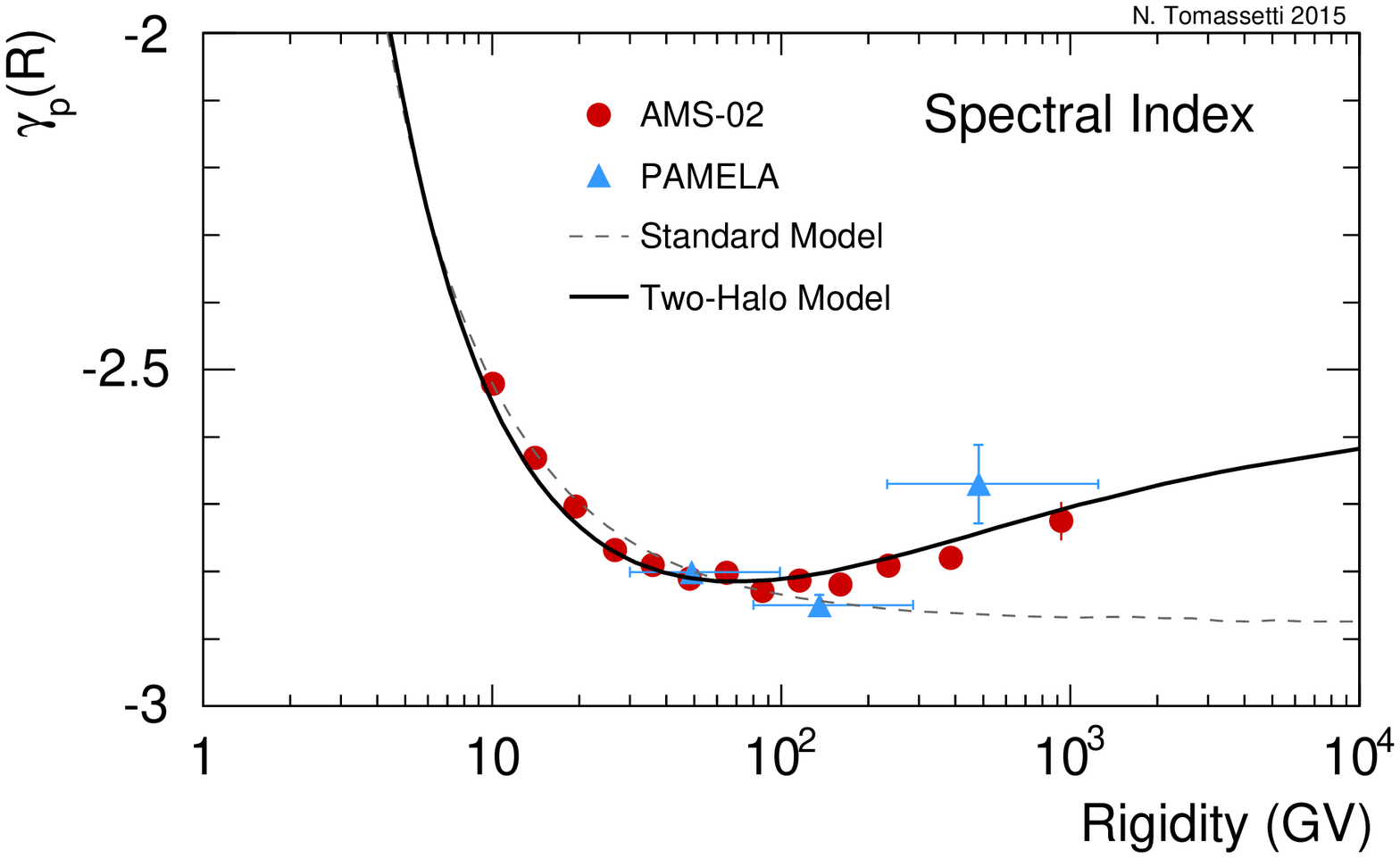}
\caption{ 
  Left: Proton energy spectrum multiplied by $E^{2.7}$. Right: rigidity dependence of the proton spectral index. 
  The model calculations are shown in comparison with the data \citep{Aguilar2015,Adriani2011,Yoon2011,Panov2009}.
  The modulation potential is $\Phi=550$\,MV.
}
\label{Fig::ccProtonSpectrum}
\end{figure}
%%%%%%%%%%%%%%%%%%%%%%%%%%%%%%%%%%%%%%%%%%%%%%%%%%%%
The proton spectrum is shown in Fig.\,\ref{Fig::ccProtonSpectrum}, for the two models, in comparison
with the data recently reported by \AMS{} \citep{Aguilar2015}.
The \AMS{} collaboration has now precisely measured the detailed variations of the proton and \He{} fluxes
at GeV--TeV energies \citep{Aguilar2015}. These data show a significantly smoother spectral hardening in comparison to that previously reported by PAMELA. 
In the figure we also compare the differential spectral index as function of rigidity, $\gamma(R)=d[\log(\phi)]/d[log(R)]$,
which is found to progressively harden at energy above $\sim$\,100\,GeV \citep{Aguilar2015}.
The results of this calculation nicely confirm the results previously obtained in our analytical derivations \citep{Tomassetti2012}.

%%%%%%%%%%%%%%%%%%%%%%%%%%%%%%%%%%%%%%%%%%%%%%%%%%%%%%%%
\section{Predictions for Secondary Cosmic-Ray Nuclei} %%%
%%%%%%%%%%%%%%%%%%%%%%%%%%%%%%%%%%%%%%%%%%%%%%%%%%%%%%%%

Secondary nuclei like boron, have steeper spectra due to diffusion. In our model, they also
experience a stronger change of slope, due to the hardening factor of their progenitor nuclei 
and their subsequent diffusion in the two propagation zones. 
%%%%%%%%%%%%% NUCLEI %%%%%%%%%%%%%%%%%%%%%%%
\begin{figure}[!t]
\includegraphics[width=0.46\textwidth]{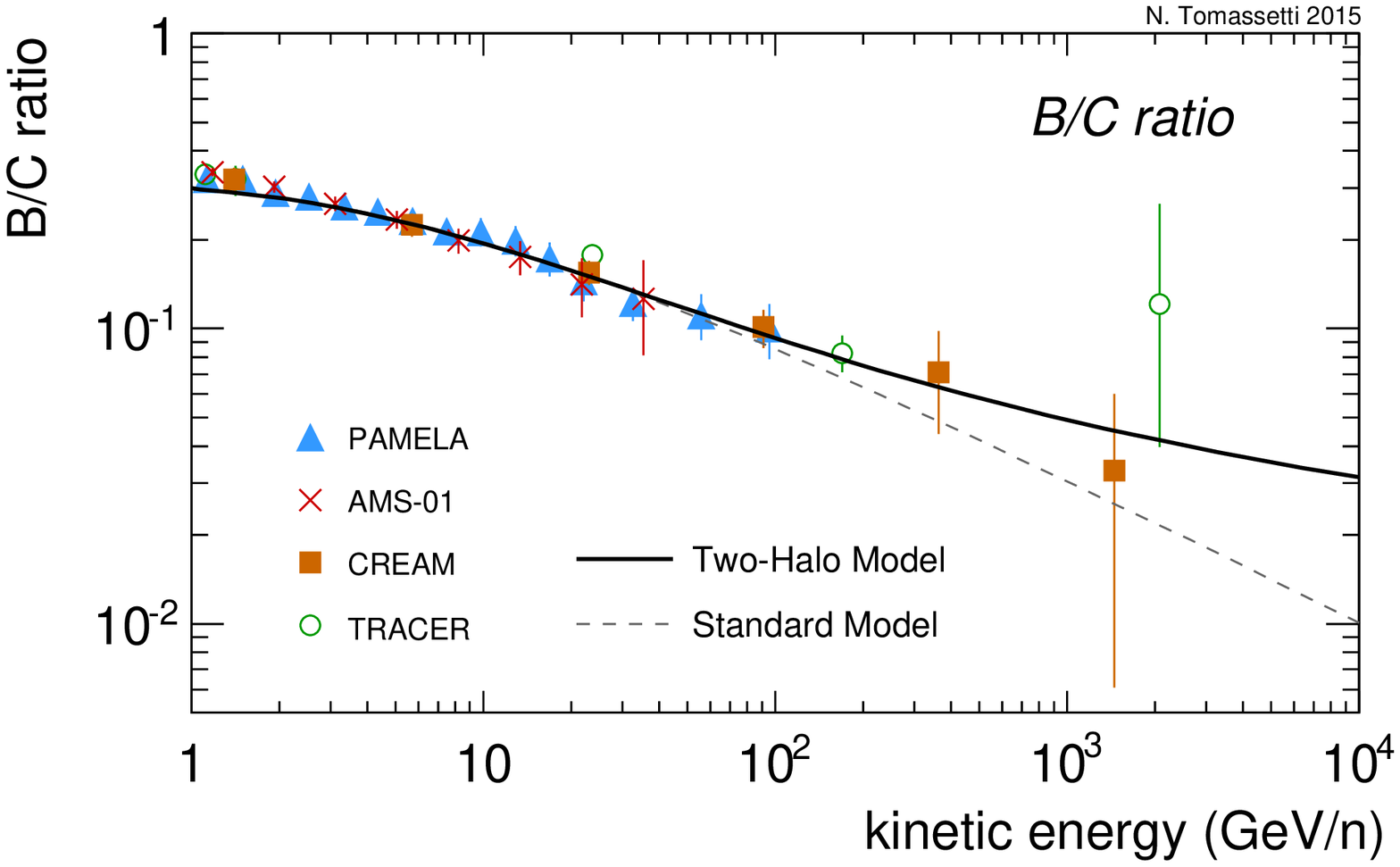}
\qquad
\includegraphics[width=0.46\textwidth]{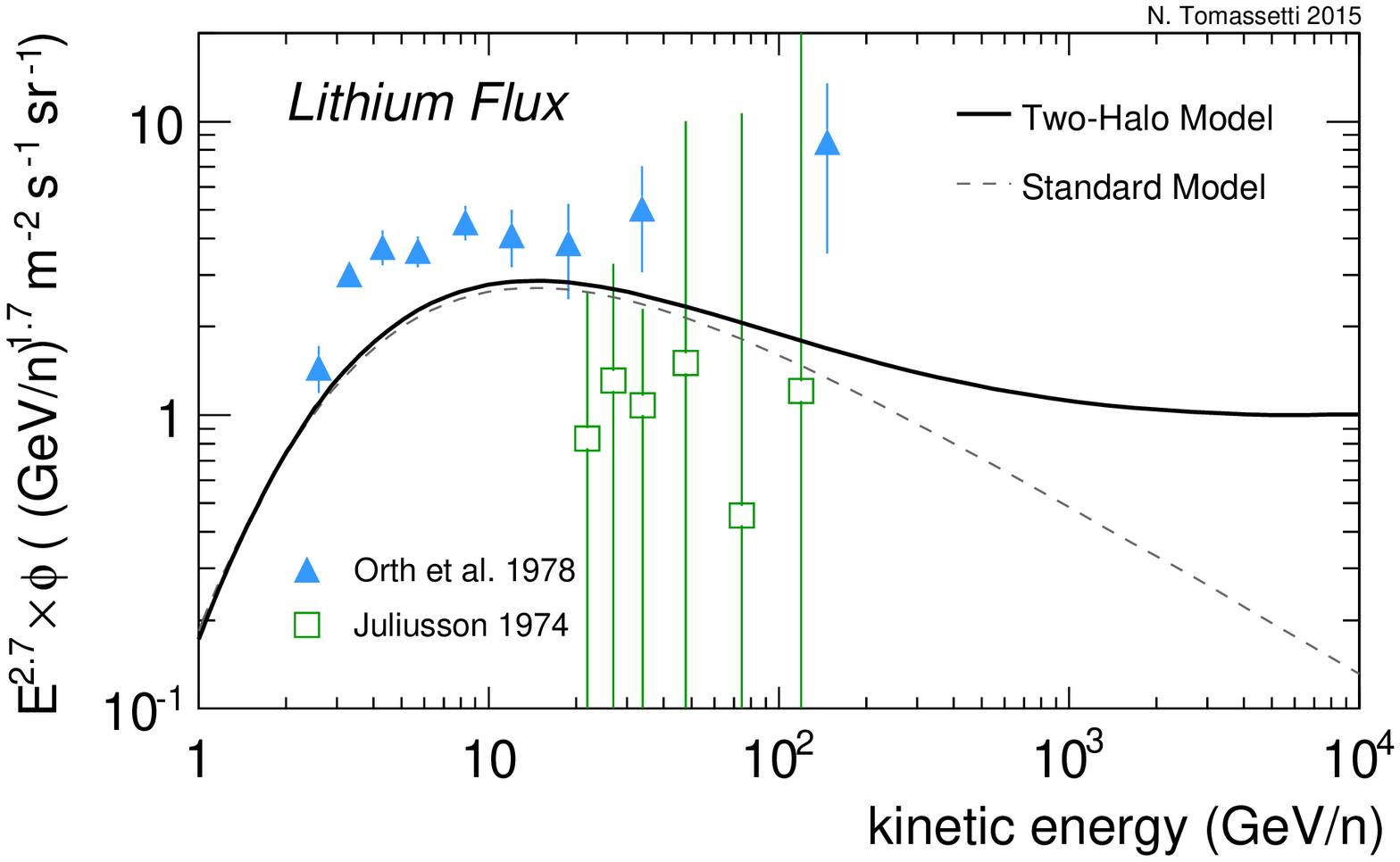}
\caption{ 
  \BC{} ratio and \Li{} spectrum. The \Li{} spectrum is multiplied by $E^{2.7}$.
  The model calculations are shown in comparison with the data \citep{Adriani2014,Ahn2010,Obermeier2012,Panov2009,Aguilar2010,Orth1978,Juliusson1974}.
  The modulation potential is $\Phi=350$\,MV.
}\label{Fig::ccNucleiSpectra}
\end{figure}
%%%%%%%%%%%%%%%%%%%%%%%%%%%%%%%%%%%%%%%%%%%%
The \BC{} ratio is therefore expected to progressively flatten at high-energy,
with an asymptotic multi-TeV dependence $\sim E^{-\delta_{i}}$ which is determined by the inner halo diffusion properties.
At low energy, our two-halo model remarkably recover the standard model behavior $\sim E^{-0.5}$. 
The existing \BC{} data support this high-energy flattening. 
The \BC{} ratio is currently being precisely measured by \AMS{} in the GeV--TeV energy region.
In the figure we also shown the Lithium spectrum. Lithium is produced by secondary reactions
of \C-\N-\Oxy{} nuclei with the ISM, like boron, but also from 
\emph{tertiary} production reactions such as \B$\rightarrow$\Li{} or \Be$\rightarrow$\Li{} which
steepen its low-energy spectrum. 
Though the existing measurements are affected by very large errors, the \Li{} flux is currently 
being measured by \AMS{} 1 to 1000 GeV/nucleon energies to a percent precision.
Measurements of \BC{} ratio and \Li{} spectrum will allow to resolutely test our scenario.

%%%%%%%%%%%%%%%%%%%%%%%%%%%%%%%%%%%%%%%%%%%%%%%%%%%%%%%%%%%%
\section{Predictions for Secondary Cosmic-Ray Antimatter} %%%
%%%%%%%%%%%%%%%%%%%%%%%%%%%%%%%%%%%%%%%%%%%%%%%%%%%%%%%%%%%%

The sharp rise of the positron fraction \epfrac{} at $\sim$10--300 GeV of energy \citep{Accardo2014,Adriani2009}, in contrast
to the standard expectations $E^{-\delta}$, may suggest the presence of extra sources of high-energy $e^{\pm}$,
including dark matter particle annihilation/decay or nearby astrophysical sources
as well as the need to reassess 
the secondary \epm{} production from CR collisions with the gas \citep{Choi2014,Blum2013,Serpico2011,Blasi2009,TomassettiDonato2015}. 
Recent \AMS{} data have triggered several speculation about a possible antiproton
excess induced by dark  matter annihilations \citep{Chen2015}.
Understanding secondary production of antimatter in CRs is of fundamental importance,
in the context of dark matter searches, in assessing \emph{both} the signal and background. 
The positron spectra, the positron fraction \epfrac, and the \pbarp{} ratio
from our model will be presented at this conference and reported in a forthcoming work \cite{Tomassetti2015thm}

%%%%%%%%%%%%%%%%%%%%%%%%%%%%%%%%%%%%%%%%%%%%%%%%%%%%%%%%%%%%%%%
\section{Connections with Anisotropy and Gamma-Ray Physics} %%%
%%%%%%%%%%%%%%%%%%%%%%%%%%%%%%%%%%%%%%%%%%%%%%%%%%%%%%%%%%%%%%%

The study on CR anisotropy in the arrival directions of CRs  has received much attention in recent years \citep{Pohl2013,Kumar2014}.
The measured CR anisotropy amplitude in at TeV--PeV is found to be nearly constant to $\lesssim$\,10$^{-3}$. 
From the standard diffusion models, the expected anisotropy exceeds the observational limits by one order of magnitude. 
To match the data, it is known that a shallower diffusivity is needed, which is in contrast to standard
model extrapolations of the \BC{} ratio. In the model considered here, this is not the case. 
Within the diffusion approximation, our model is able to reconcile the anisotropy amplitude \emph{and} the \BC{} ratio.

Concerning $\gamma$-ray physics, we note that the \emph{Fermi-LAT} collaboration has reported
a $\gamma$--ray spectral flattening of the  spectra in the inner Galaxy in comparison with
the outer Galaxy or the halo \citep{Ackermann2012}. These observations 
find a natural explanation in spatial changes of the CR diffusion properties \citep{ErlykinWolfendale2002,Tomassetti2012}.
%%This dependence has been considered in \citep{ErlykinWolfendale2012} and recently tested using \Dragon{} \citep{Gaggiero2015}. 

%%%%%%%%%%%%%%%%%%%%%%%%%
\section{Conclusions} %%%
%%%%%%%%%%%%%%%%%%%%%%%%%

Several properties observed in the CR spectrum can found a natural explanation 
under a diffusive propagation scenario characterized by a spatial change of the CR
transport properties in the Galaxy. 
By change of transport properties, we refer to a different energy dependence of the
CR diffusion in different region of the Galaxy. 
An important ingredient is that the diffusion coefficient has to be a non-separable function
of energy and space coordinates, in contrast to standard diffusion models that assume homogeneous 
diffusivity --or space-energy separability-- in the whole Galactic halo.
A consequence of this scenario is the explanation of the spectral hardening 
observed in primary CRs, and in particular in the new \AMS{} proton data.
Remarkably, this scenario is able to simultaneously accounts for the low-energy steepness 
of the \BC{} ratio \emph{and} for the high-energy requirements of CR anisotropy studies.
Important consequences of our scenario deals with the secondary antimatter spectra.
More quantitative studies will be presented upon the release of the \AMS{} data on \Li, \BC{} and \pbarp.

%%%%%%%%%%%%%%%%%%%%%%%%%%%%%%%%%%%%%%%%%%%%%%%%%%%%%%%%

\end{document}